\documentclass[12pt]{article}
\usepackage{epsfig}
\def\beq{\begin{equation}}
\def\eeq{\end{equation}}

\def\be{\begin{equation}}
\def\bea{\begin{eqnarray}}
\def\ee{\end{equation}}
\def\eea{\end{eqnarray}}

\def\eqref#1{(\ref{#1})}

\newcommand{\al}{\alpha}

\newcommand{\dl}{\delta}

\begin{document}

%----------------------------------------------------------------------
%                     T I T L E
%----------------------------------------------------------------------

\begin{titlepage}
\begin{centering}

{\Large\bf ${\cal N}=(1,1)$ super Yang--Mills theory in 1+1
dimensions at finite temperature}\\
\vspace*{1.5cm}

{\bf John R.~Hiller}
\vspace*{0.5cm}

{\sl Department of Physics \\
University of Minnesota Duluth\\
Duluth MN  55812}
\vspace*{0.5cm}

{\bf  Yiannis Proestos, Stephen Pinsky, and Nathan Salwen}
\vspace*{0.5cm}

{\sl Department of Physics \\
Ohio State University\\
Columbus OH 43210}

\vspace*{1cm}

%----------------------------------------------------------------------%
%                       A B S T R A C T
%----------------------------------------------------------------------%

\vspace*{1cm}

\begin{abstract}
We present a formulation of ${\cal N}=(1,1)$ super Yang--Mills theory
in 1+1 dimensions at finite temperature. The partition function is
constructed by finding a numerical approximation to the entire
spectrum. We solve numerically for the spectrum using Supersymmetric
Discrete Light-Cone Quantization (SDLCQ) in the large-$N_c$
approximation and calculate the density of states.  We find that the
density of states grows exponentially and the theory has a Hagedorn
temperature, which we extract. We find that the Hagedorn temperature
at infinite resolution is slightly less than one in units of
$\sqrt{g^{2} N_{\rm c}/\pi}$.  We use the density of states to also
calculate a standard set of thermodynamic functions below the
Hagedorn temperature.  In this temperature range, we find that the
thermodynamics is dominated by the massless states of the theory.
\end{abstract}
\end{centering}

%\noindent
%PACS number(s):
\vfill

\end{titlepage}
\newpage

%----------------------------------------------------------------------%
%                       Introduction
%----------------------------------------------------------------------%
%%%%%%%%%%%%%%%%%%%%%%%%%%%%%%%%%%%%%%%%%%%%%%%%%%%%%%%%%%%%%%%%%%
\section{Introduction}
%%%%%%%%%%%%%%%%%%%%%%%%%%%%%%%%%%%%%%%%%%%%%%%%%%%%%%%%%%%%%%%%%%
In ${\cal N}=4$ super Yang--Mills theory at large $N_c$, there is a
known mismatch between weak and strong coupling by a factor of 3/4
in the free energy~\cite{Gubser:1996de}. The weak-coupling result
is calculable in perturbation theory; the strong-coupling result
comes from black-hole thermodynamics. It would be interesting to
be able to directly solve this theory at all couplings and see the
transition between the weak-coupling and strong-coupling
limits~\cite{Li:1998kd}.
Analytically this is generally not possible, although there have
been a number of early discussions of methods for
finite-temperature solutions to supersymmetric quantum field
theory~\cite{Das:rx}.  We will instead consider a numerical method
based on Supersymmetric Discrete Light-Cone Quantization
(SDLCQ)~\cite{sakai,Lunin:1999ib}, which preserves the
supersymmetry exactly.  Currently this is the only method
available for numerically solving strongly coupled super
Yang--Mills theories.  Conventional lattice methods have
difficulty with supersymmetric theories because of the asymmetric
way that fermions and bosons are treated, and
progress~\cite{lattice} in supersymmetric lattice gauge theory has
been relatively slow.

Given that SDLCQ makes use of light-cone coordinates, with
$x^+=(x^0+x^3)/\sqrt{2}$ the time variable and
$p^-=(p^0-p^3)/\sqrt{2}$ the energy, we must take some care in
defining thermodynamic quantities. The seemingly natural
choice~\cite{Brodsky:2001bx} of $e^{-\beta_{_{\rm LC}}p^-}$ as the
partition function has been shown by Alves and
Das~\cite{Alves:2002tx} to lead to singular results for well known
quantities that are finite in the equal-time approach. They argue
that using $e^{- \beta_{_{\rm LC}} p^-}$ for the partition
function implies that the physical system is in contact with a
heat bath that has been boosted to the light-cone frame and that
this is not equivalent to the physics of a system in contact with
a heat bath at rest.

A more direct way to see this is that, since the light-cone
momentum $p^+=(p^0+p^3)/\sqrt{2}$ is conserved, the partition
function must include it in the form $Z=e^{-\beta_{_{\rm LC}}(p^-
+ \mu p^+)}$, with $\mu$ its chemical potential.  The
interpretation of the chemical potential is that of a rotation of
the quantization axis. Thus $\mu=1$ corresponds to quantization in
an equal-time frame, where the heat bath is at rest and the inverse
temperature
$\beta=\sqrt{2}\beta_{_{\rm LC}}$, and $\mu \neq 1$
corresponds to quantization in a boosted frame where the heat bath
is not at rest. Thus $\mu$ corresponds to a continuous rotation of
the axis of quantization, and $\mu=0$ would correspond to rotation
all the way to the light-cone frame. A rotation from an equal-time
frame to the light-cone frame is not a Lorentz transformation. It
is known that such a transformation can give rise to singular
results for physical quantities. This appears to be consistent
with the results found in~\cite{Alves:2002tx}. A number of related
issues have been extensively discussed by
Weldon~\cite{Weldon:aq}. The method has also
recently been applied to the Nambu--Jona-Lasinio
model~\cite{Beyer:2003qb}.

These difficulties are avoided if we compute the equal-time
partition function $Z=e^{-\beta p^0}$, as was proposed much
earlier by Elser and Kalloniatis~\cite{Elser:1996tq}.  The
computation may, of course, still use light-cone coordinates.
Elser and Kalloniatis did this with ordinary
DLCQ~\cite{pb85,bpp98} as the numerical approximation to
(1+1)-dimensional quantum electrodynamics.  Here we will follow a
similar approach using SDLCQ to calculate the spectrum of ${\cal
N}=(1,1)$ super Yang--Mills theory in 1+1
dimensions~\cite{Antonuccio:1998kz}. Though this
calculation is done in 1+1 dimensions, it is known that SDLCQ can
be extended in a straightforward manner to higher
dimensions~\cite{Antonuccio:1999zu, Haney:2000tk,hpt2001}.

We have discussed the SDLCQ numerical method in a number of other
places, and we will not present a detailed discussion of the method
here; for a review, see~\cite{Lunin:1999ib}. For those familiar
with DLCQ~\cite{pb85,bpp98}, it suffices to say that SDLCQ is
similar; both have discrete momenta and cutoffs in momentum space,
$x^- \in [-L,L]$.
In 1+1 dimensions the discretization is specified by a single
integer $K=(L/\pi) P^+$, the resolution, such that longitudinal
momentum fractions are integer multiples of $1/K$. However, SDLCQ is
formulated in such a way that the theory is also exactly
supersymmetric.  Exact supersymmetry brings a number of very
important numerical advantages to the method; in particular,
theories with enough supersymmetry are finite. We have also seen
greatly improved numerical convergence in this approach.

In Sec.~\ref{sec:Super} we briefly review super Yang--Mills theory
in 1+1 dimensions.  The discussion in
Sec.~\ref{sec:density} describes the method we use to extract the
density of states from the numerical spectrum.
The calculation of the density of states is presented in
Sec.~\ref{sec:results}.  We fit the data to smooth analytical
functions, and we find that the theory has a Hagedorn temperature
$T_H$~\cite{Hagedorn_NC_65_68}, which we calculate.
In Sec.~\ref{sec:Basic formulae}, we use the analytic fit to the
density of states to calculate the free energy, the energy, and the
specific heat, up to the Hagedorn temperature.
Section~\ref{sec:discussion} contains a discussion of
our results and the prospects for future work using these methods.

%%%%%%%%%%%%%%%%%%%%%%%%%%%%%%%%%%%%%%%%%%%%%%%%%%%%%%%%%%%%%%%%%%
\section{Super Yang--Mills theory} \label{sec:Super}
%%%%%%%%%%%%%%%%%%%%%%%%%%%%%%%%%%%%%%%%%%%%%%%%%%%%%%%%%%%%%%%%%%
We will start by providing a brief review of ${\cal N}=(1,1)$
supersymmetric Yang--Mills in 1+1 dimensions. The Lagrangian of
this theory is
\begin{equation}
{\cal L}
 ={\rm Tr}\left(-\frac{1}{4}F_{\mu\nu}F^{\mu\nu}
      +i\bar{\Psi}\gamma_{\mu}D^{\mu}\Psi\right).\label{Lagrangian}
\end{equation}
The two components of the spinor $\Psi=2^{-1/4}({\psi \atop
\chi})$ are in the adjoint representation, and we will work in the
large-$N_c$ limit.  The field strength and the covariant
derivative are
$F_{\mu\nu}=\partial_{\mu}A_{\nu}-\partial_{\nu}A_{\mu}
+ig[A_{\mu},A_{\nu}]$ and $D_{\mu}=\partial_{\mu}+ig[A_{\mu},~]$.
The most straightforward way to formulate the theory in 1+1
dimensions is to start with the theory in 2+1 dimensions and then
simply dimensionally reduce to 1+1 dimensions by setting
$\phi=A_2$ and $\partial_2 \rightarrow 0$ for all fields.  In the
light-cone gauge, $A^+=0$, we find
\begin{equation}
Q^-=2^{3/4}g\int dx^- {\rm Tr}\left(i[\phi,\partial_-\phi]\frac{1}{\partial_-}\psi
         +2\psi\psi\frac{1}{\partial_-}\psi\right) .
\end{equation}
The mode expansions in two dimensions are
\begin{eqnarray}
\phi_{ij}(0,x^-) &=& \frac{1}{\sqrt{2\pi}} \int_0^\infty
         \frac{dk^+}{\sqrt{2k^+}}\left[
         a_{ij}(k^+)e^{-{\rm i}k^+x^-}+
         a^\dagger_{ji}(k^+)e^{{\rm i}k^+x^-}\right] ,
\nonumber\\
\psi_{ij}(0,x^-) &=&\frac{1}{2\sqrt{\pi}}\int_0^\infty
         dk^+\left[b_{ij}(k^+)e^{-{\rm i}k^+x^-}+
         b^\dagger_{ji}(k^+)e^{{\rm i}k^+x^-}\right] .
\end{eqnarray}
To obtain the spectrum, we solve the mass eigenvalue problem
\begin{equation}
2P^+P^-|\varphi\rangle=\sqrt{2}P^+(Q^-)^2|\varphi\rangle
     =M^2|\varphi\rangle .
\label{EVP}
\end{equation}

This theory has two discrete symmetries, besides supersymmetry, that
we use to reduce the size of the Hamiltonian matrix we have to
calculate. $S$-symmetry, which is associated with the orientation
of the large-$N_c$ string of partons in a state~\cite{kut93},
gives a sign when the color indices are permuted
\begin{equation}\label{Z2-S}
S : a_{ij}(k)\rightarrow -a_{ji}(k) , \qquad
      b_{ij}(k)\rightarrow -b_{ji}(k) .
\end{equation}
$P$-symmetry is what remains of parity in the $x^2$ direction
after dimensional reduction
\begin{equation}\label{Z2-P}
P : a_{ij}(k)\rightarrow -a_{ij}(k), \qquad
      b_{ij}(k)\rightarrow b_{ij}(k).
\end{equation}
All of our states can be labeled by the $P$ and $S$ sector in
which they appear.

We convert the mass eigenvalue problem $2P^+P^-|M\rangle = M^2
|M\rangle$ to a matrix eigenvalue problem by introducing a
discrete basis where $P^+$ is diagonal. We will always state $M^2$
in units of $g^2 N_c/\pi$. In SDLCQ the discrete basis is introduced
by first discretizing the supercharge $Q^-$ and then constructing
$P^-$ from the square of the supercharge: $P^- =
(Q^-)^2/\sqrt{2}$.  To discretize the theory, we impose periodic
boundary conditions on the boson and fermion fields alike and
obtain an expansion of the fields with discrete momentum modes. We
define the discrete longitudinal momenta $k^+$ as fractions
$nP^+/K$ of the total longitudinal momentum $P^+$, where $K$ is an
integer that determines the resolution of the discretization and
is known in DLCQ as the harmonic resolution~\cite{pb85}. Because
light-cone longitudinal momenta are always positive, $K$ and each
$n$ are positive integers; the number of constituents is then
bounded by $K$.   The continuum limit is recovered by taking the
limit $K \rightarrow \infty$.

In constructing the discrete approximation, we drop the
longitudinal zero-momentum mode.  For some discussion of dynamical
and constrained zero modes, see the review~\cite{bpp98} and
previous work~\cite{alpt98}. Inclusion of these modes would be
ideal, but the techniques required to include them in a numerical
calculation have proved to be difficult to develop, particularly
because of nonlinearities.   For DLCQ calculations that can be
compared with exact solutions, the exclusion of zero modes does
not affect the massive spectrum~\cite{bpp98}. In scalar theories
it has been known for some time that constrained zero modes can
give rise to dynamical symmetry breaking~\cite{bpp98}, and work
continues on the role of zero modes and near zero modes in these
theories~\cite{Rozowsky:2000gy,Salmons:2002xg,Heinzl:2003jy,%
Harindranath:ud,Chakrabarti:2003ha}.

%%%%%%%%%%%%%%%%%%%%%%%%%%%%%%%%%%%%%%%%%%%%%%%%%%%%%%%%%%%%%%%%%%%%%%%%%%%%
%
\section{Density of states}
\label{sec:density}
%%%%%%%%%%%%%%%%%%%%%%%%%%%%%%%%%%%%%%%%%%%%%%%%%%%%%%%%%%%%%%%%%%%%%%%%%%%%
%%
%
%
The thermodynamic functions will be written as sums over the spectrum
$\{M_n\}$ of the theory.
The most convenient way to calculate such sums is
to represent each sum as an integral over a density of states
$\rho(M^2)$,
\begin{equation}
\sum_{n=1}^{\infty} \rightarrow \int \rho(M^2) dM^2 .
\label{f10}
\end{equation}
{}From our numerical solutions we can approximate the density of
states by a continuous function. The remaining integrals in the
thermodynamic functions are then done by standard numerical
integration techniques, which are fast and convenient.

We can look at the density for a series of increasing resolutions
$K$ in the SDLCQ approximation and thereby discuss the convergence
of the density in the limit $K\rightarrow \infty$.  The maximum
mass that we can reach in the SDLCQ approximation increases as we
increase the resolution.  We
report results for $11 \leq K\leq 17$.

Convenient functions to extract from the spectral
data~\cite{Lunin:2000im} are the cumulative distribution function
(CDF) $N(M^2,K)$ and the normalized cumulative
distribution function (NCDF) $f(M^2,K,M^2_r)$. The CDF is the number of
massive states at or below $M^2$ at resolution $K$, and the NCDF is
this number divided by the total number of massive states below
an arbitrary normalization point $M^2_r$, again at resolution $K$:
\begin{equation}
f(M^2,K,M^2_r)   = \frac{N(M^2, K)}{N(M^2_r,K)}.
\label{ffunction}
\end{equation}
The function $f$ turns out to be very smooth and can be fit by a single
smooth analytic form. By definition, the density of states is given by
\begin{equation}
\rho(M^2,K) = \frac{dN\left(M^2,K\right)}{dM^2} .
\end{equation}
It is also convenient to define a normalized density of
states~\cite{Lunin:2000im}
\begin{equation}
\tilde{\rho}(M^2,K,M^2_r)=\frac{df\left(M^2,K,M^2_r\right)}{dM^2}.
\end{equation}
It is well known that
if the density of states grows exponentially with the mass of the state,
\begin{equation}
\rho(M^2)\sim {\rm{exp}}(M/T_{\rm{H}}), \label{f10a}
\end{equation}
the theory will have a Hagedorn temperature,
$T_{\rm{H}}$~\cite{Hagedorn_NC_65_68}. Above the temperature $T_H$,
the thermodynamic integrals diverge.

%%%%%%%%%%%%%%%%%%%%%%%%%%%%%%%%%%%%%%%%%%%%%%%%%%%%%%%%%%%%%%%%%%%%%%%%%%%%
%%
\section{Numerical results for the density of states}
\label{sec:results}
%%%%%%%%%%%%%%%%%%%%%%%%%%%%%%%%%%%%%%%%%%%%%%%%%%%%%%%%%%%%%%%%%%%%%%%%%%%%
%%
The numerical results presented in this section are the first from our new
code, which was rewritten to run on clusters. Most of these results
were produced on our six-processor development cluster. While this
cluster was sufficient for this problem, we expect to be able to
handle larger problems by moving to larger clusters. In fact, it is
now so easy to generate the Hamiltonian up to resolution $K=17$,
that we only used one node in our cluster for that purpose. What
made this calculation challenging numerically was that we needed
to extract a large number of eigenvalues.  For the largest values
of the resolution, this was done with a specially tuned Lanczos
diagonalization code based on the techniques of Cullum and
Willoughby~\cite{Cullum}. We should note that, prior to the development
of the new matrix-generation code, the highest resolution results
presented for this theory were for resolution $K=10$~\cite{Antonuccio:1998kz}.
Here we will present only resolutions $K\geq 11$. All our earlier
results are now trivially reproduced.

\begin{figure}[ht]
\begin{tabular}{cc}
\psfig{figure=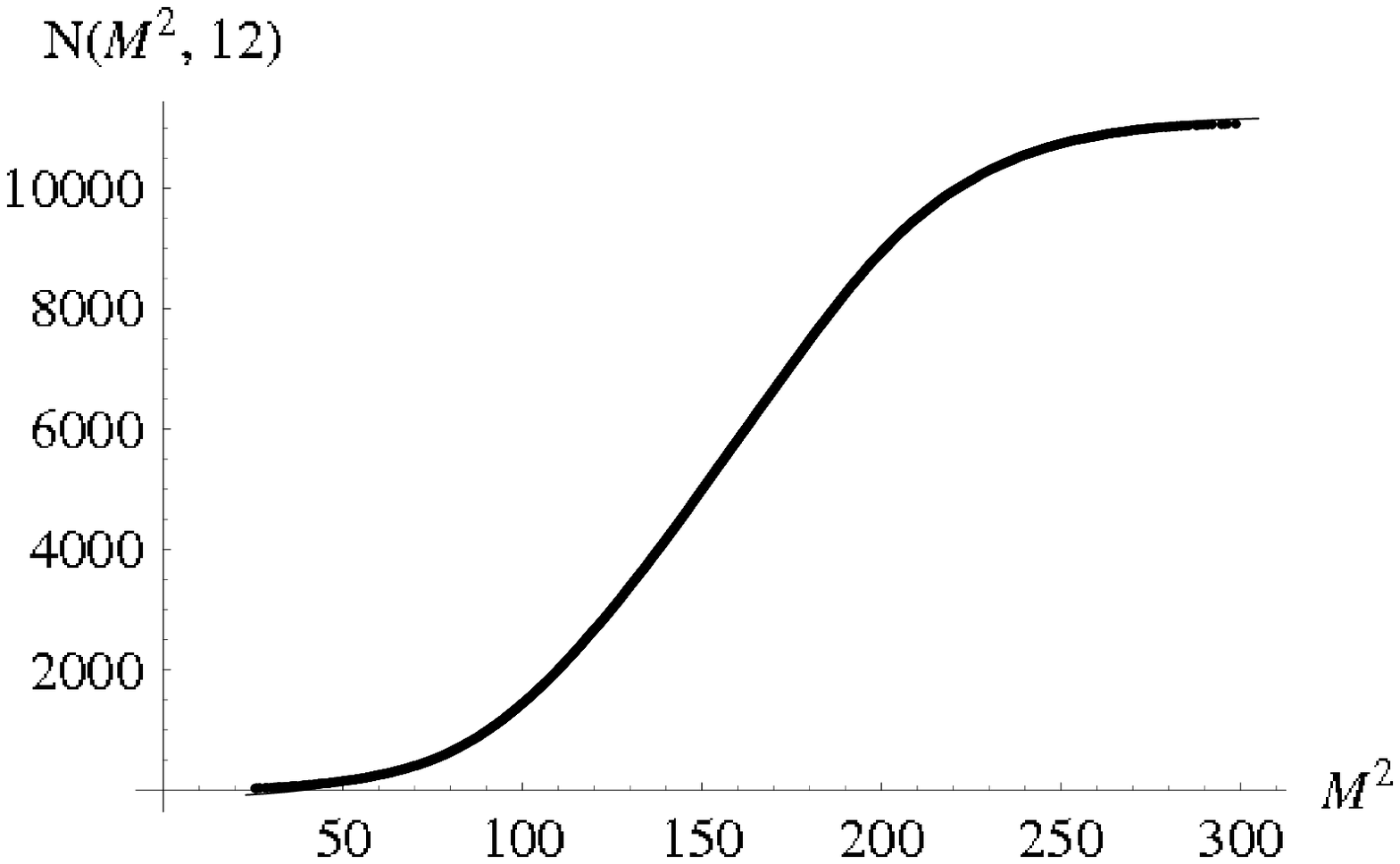,width=6.8cm,angle=0}&
\psfig{figure=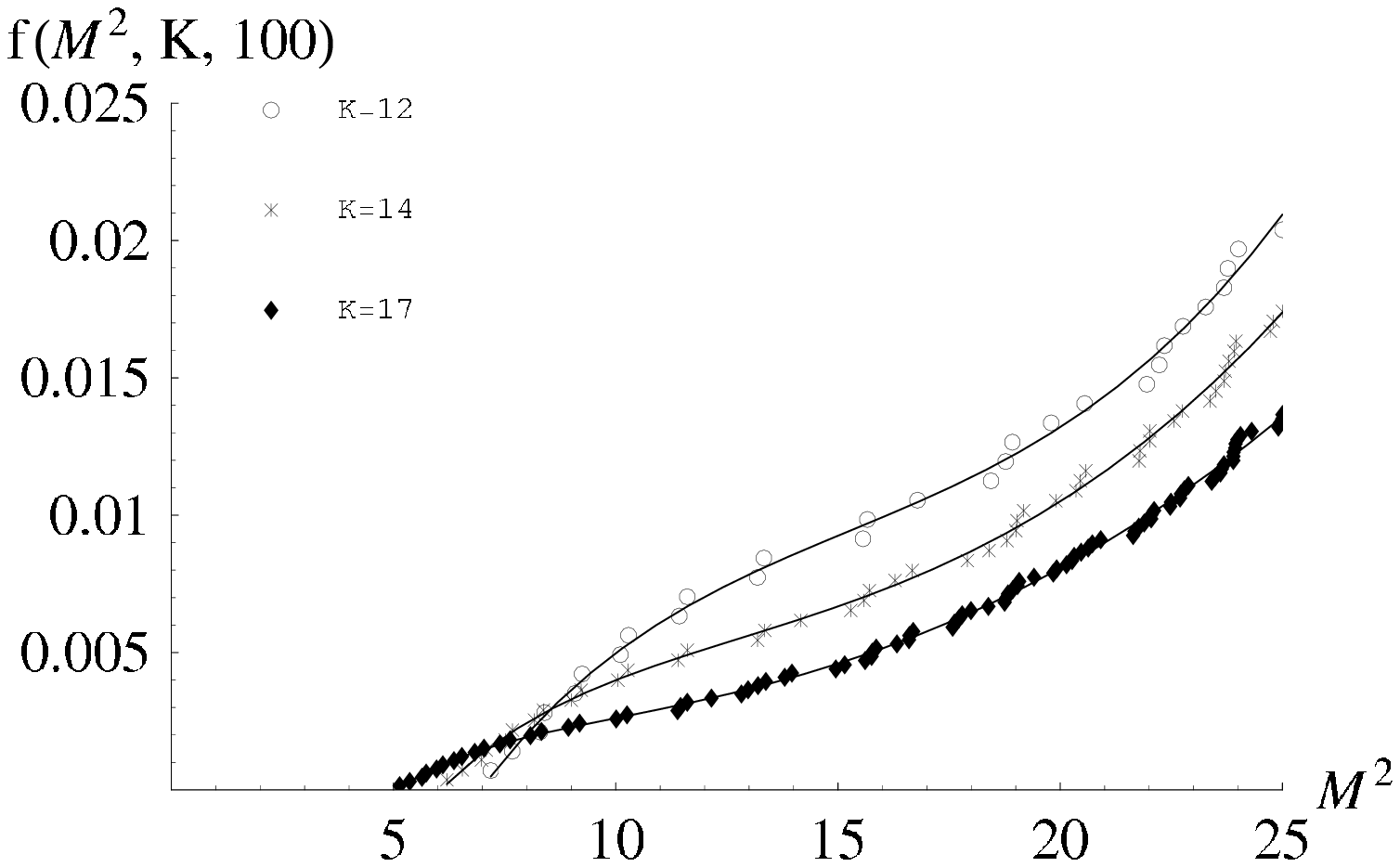,width=6.8cm,angle=0}\\
(a) & (b)
\end{tabular}
\caption{The distribution functions (a) $N(M^2,K)$ for $K=12$
and (b) $f(M^2,K,100)$ for $K=12$, 14, and 17, as functions of
$M^2$ in units of $g^2 N_c/\pi$.  In (a) we also show
the error-function fit $a \,{\rm erf}(b (x-c))+d$, and in (b),
a polynomial fit.}
\label{NtotalK12}
\end{figure}

In the SDLCQ approximation, the portion of the spectrum that we can
see at any finite resolution will naturally be cut off. As
we approach the cutoff, the approximation limits the
number of states that are available and distorts the density of
states. In Fig.~\ref{NtotalK12}a, we present the data for the CDF at
resolution
$K=12$, where we can diagonalize the entire Hamiltonian. We clearly
see evidence of this distortion.
By the midpoint, the cutoff is already diminishing the number of states
available in the approximation. Interestingly, we can find a fit to
this data with a simple universal function of the form
$a\,{\rm erf}(b (x-c))+d$. Clearly the fit shown is excellent;
the fit is so good that one cannot separately see the data and the
fitted curve on this scale.

At low masses, there is a mass gap, which has been discussed
elsewhere~\cite{Lunin:1999ib}. The mass gap closes linearly with $1/K$.
For very small values of
$M^2$, the decrease of the mass gap with increasing resolution adversely
affects the quality of the universal fit, and it is convenient in that
region to improve the quality of the fit by using  a polynomial
function of the form
\begin{equation}
f(x,K,100)=\sum_{p=0}^{p(K)}\al_p \, x^{\dl_K +p}\Theta (x-x_{\rm min}(K)).
\end{equation}
In Fig.~\ref{NtotalK12}b we show the fit to the
NCDF for some representative values of the resolution $K$. We have only
shown the data at resolutions $K=12$, 14, and 17 to keep the figure
uncluttered. One can see how the endpoints tend to lower
values as the mass gap closes with increasing $K$.

\begin{figure}
\begin{tabular}{cc}
\psfig{figure=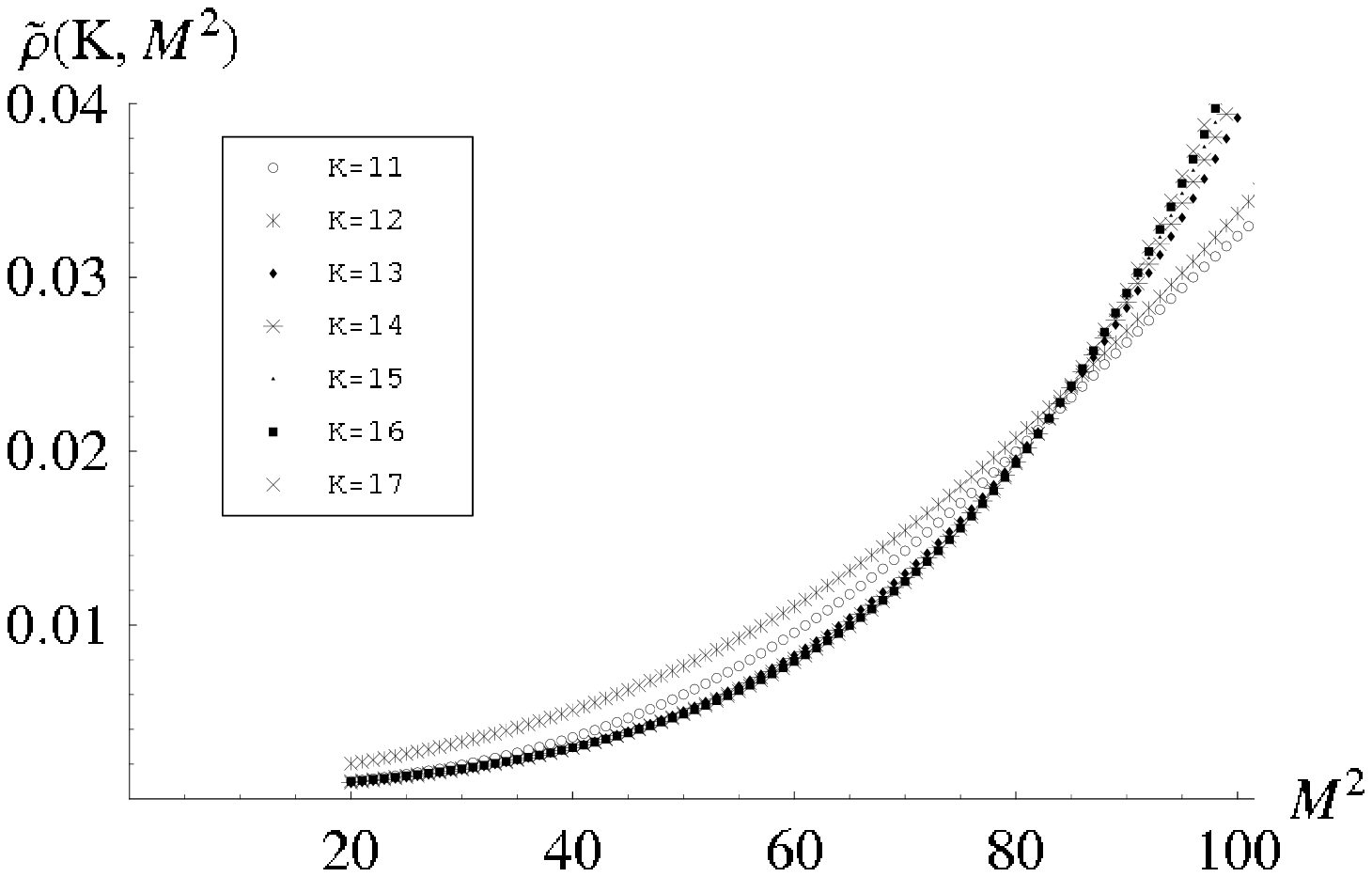,width=6.8cm,angle=0}&
\psfig{figure=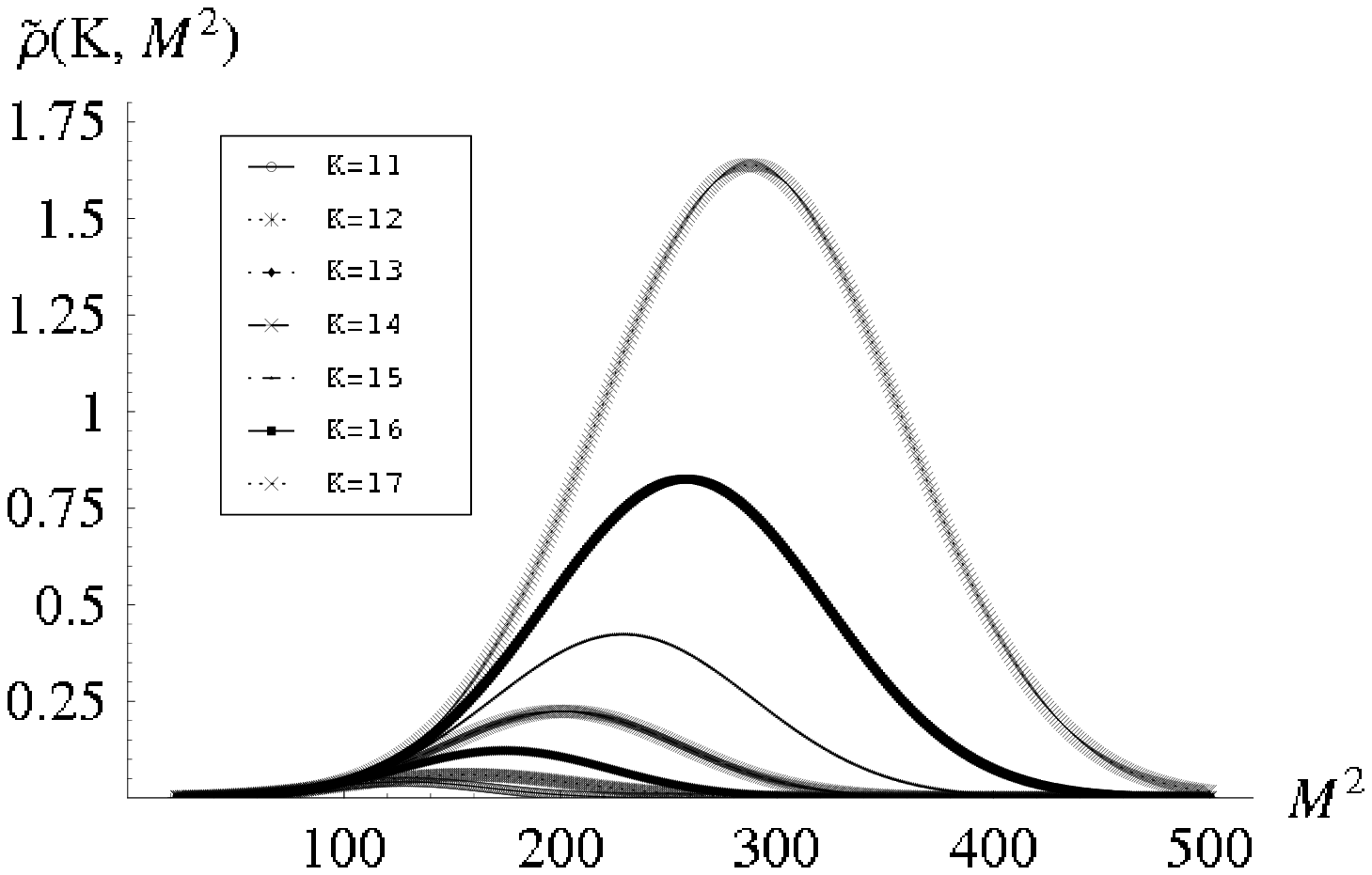,width=6.8cm,angle=0}\\
(a) & (b)
\end{tabular}
\caption{The normalized density of states (a) for $M^2 \leq 100$ and
(b) extrapolated to all masses.
The data points are a convenient way of displaying the continuous
functions calculated from the fits to the CDF.}
\label{data1}
\end{figure}
At larger values of $K$ it is difficult to completely diagonalize the
entire Hamiltonian. We have limited ourselves to states with $M^2
\leq 100$. However, once we know the universal form of the function
that fits the CDF, we can fit just the region $M^2 \leq 100$ and
extrapolate to all masses. At large $M^2$, the CDF approaches the
total number of bound states. The total number of states in the SDLCQ
approximation at any resolution, and, in any symmetry sector in the
large-$N_c$ approximation, is exactly calculable; the general
results will be discussed elsewhere. We use this asymptotic value of
the CDF, in addition to the behavior for $M^2
\leq 100$, in making the fits. We have done this at all resolutions up
to $K=17$. In Fig.~\ref{data1}a we show the normalized density of states
calculated from the NCDF, for $M^2 \leq 100$. In Fig.~\ref{data1}b we
show the normalized density of states for $11 \leq K \leq 17$,
extrapolated to the full range of masses.
\begin{figure}[ht]
\begin{tabular}{cc}
\psfig{figure=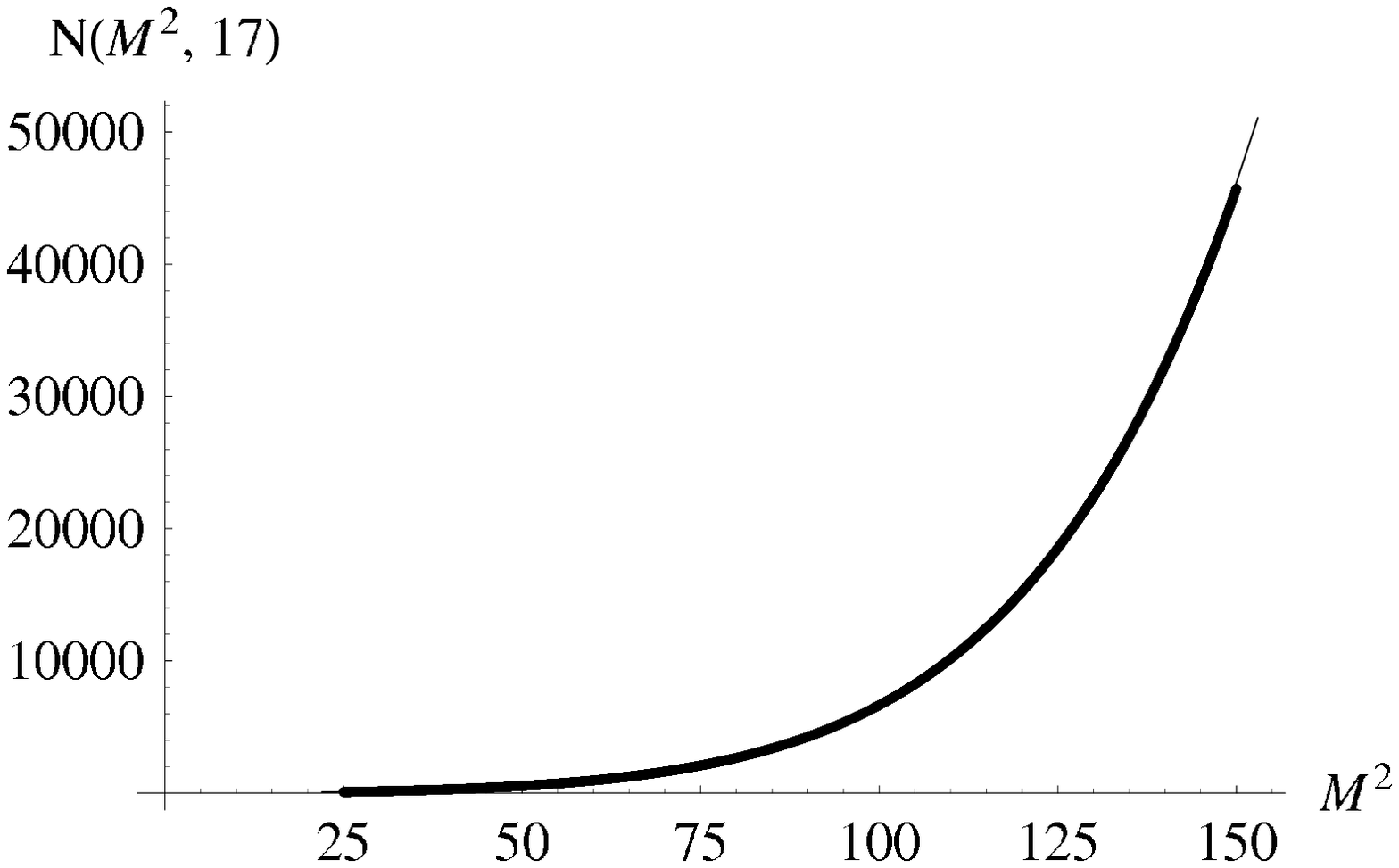,width=6.8cm,angle=0}&
\psfig{figure=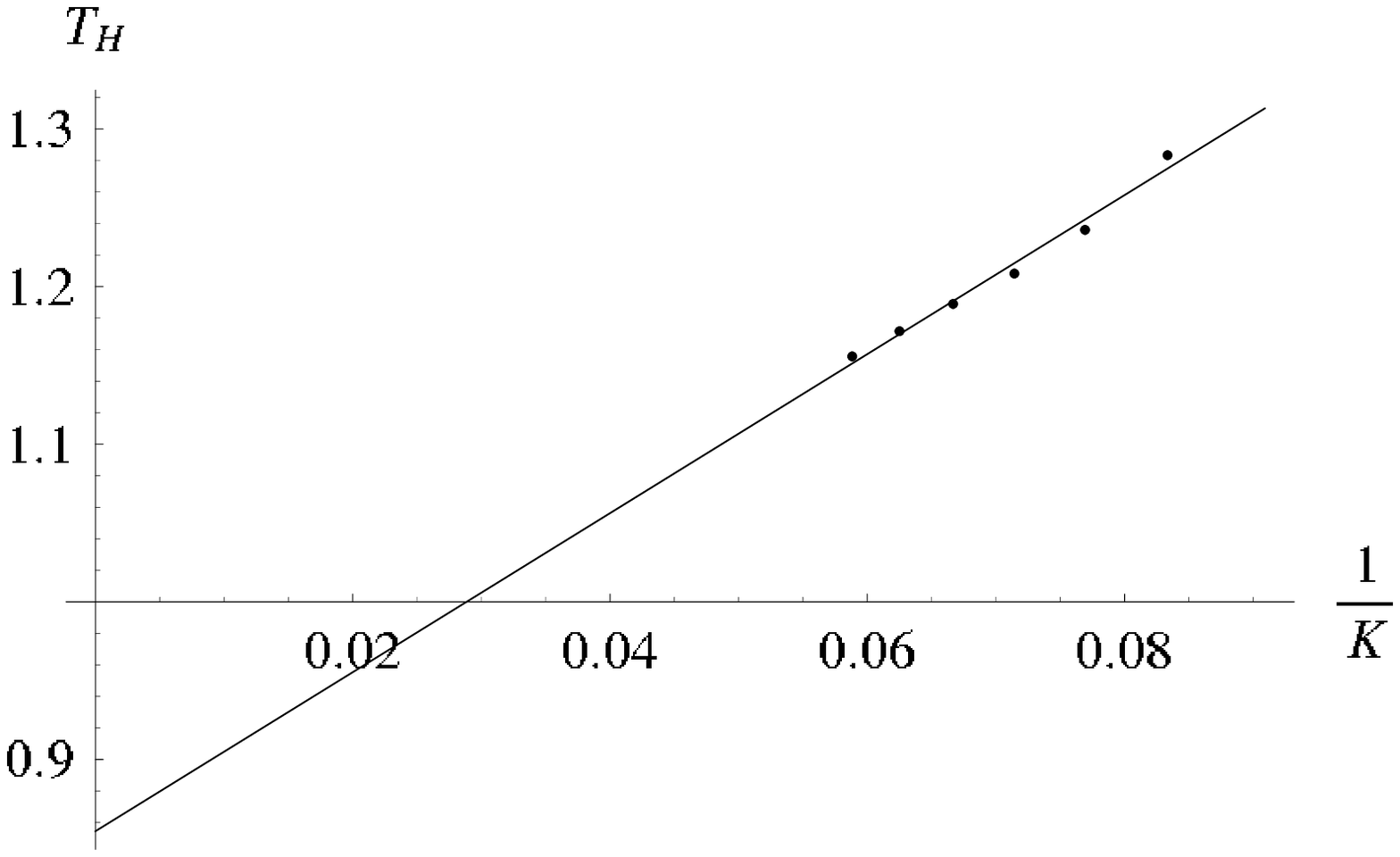,width=6.8cm,angle=0}\\
(a) & (b)
\end{tabular}
\caption{The Hagedorn temperature $T_H$, as obtained from
exponential fits, $\al \,{\rm{exp}}\left(\frac{M}{T_{\rm H}}\right) -\al$,
to the universal
fits to $N(M^2,K)$, as shown in (a) for $K=17$, and (b) a linear plot
against $1/K$ for fits in the range $11 \leq K \leq 17$.}
\label{Hagedorn}
\end{figure}

Inspecting these curves, we see that on the up-slope part of the
density of states, where we believe our numerical approximation is a
valid approximation to the actual density of states, the shape appears
exponential. As we go to larger and larger values of the resolution $K$,
the size of this region grows. This suggests that the density of
states ultimately becomes simply an exponential, and, therefore, this
theory has a Hagedorn temperature. To find the Hagedorn temperature, we
fit the NCDF in this up-slope region with a function of the form
\begin{equation}
f(M^2,K,100)=\al \, {\rm{exp}}\left(\frac{M}{T_{\rm
H}}\right) -\al. \label{fitHagedorn}
\end{equation}
In Fig.~\ref{Hagedorn}b we plot
$ T_H$ against $1/K$. This yields a good linear fit, which we
extrapolate to infinite resolution.
We find that the Hagedorn temperature at infinite resolution is
slightly less than one in units of $\sqrt{g^{2} N_{\rm c}/\pi}$.
This value serves as a limiting temperature for the region of validity
in the calculation of thermodynamic quantities.

%%%%%%%%%%%%%%%%%%%%%%%%%%%%%%%%%%%%%%%%%%%%%%%%%%%%%%%%%%%%%%%%%%
\section{Finite temperature in 1+1 dimensions}
\label{sec:Basic formulae}
%%%%%%%%%%%%%%%%%%%%%%%%%%%%%%%%%%%%%%%%%%%%%%%%%%%%%%%%%%%%%%%%%
In the large-$N_c$ approximation, the numerical solution of a
theory is a set of non-interacting bound states. Therefore, the
thermodynamics of such supersymmetric theories is simply the
thermodynamics of a gas of a large number of species of degenerate
bosons and fermions. In
principle, one could go beyond the calculation of the standard set
of the thermodynamic functions and calculate a variety of matrix
elements. These calculations would require the wave functions of
the bound states, which can be calculated as part of the SDLCQ
calculation. We will, however, not exploit this detailed information
here. We will focus on the calculation of standard thermodynamic
quantities that can be obtained from the density of states.
The light cone plays no role beyond the calculation of the density;
the thermodynamics is that of a system at rest.

Let us now briefly review the thermodynamics of free bosons and
fermions. We assume that our system has constant volume $V$ and is in
contact with a heat bath of constant temperature $T$. The free
energy in units with $k_B=1$ is
\begin{equation}
{\cal{F}}(T,V)=-T\ln Z. \label{f1}
\end{equation}
The contribution of a single bosonic oscillator to the free energy
$F_{B}$ is
\begin{equation}
F_{B}=2VT\int_{0}^{\infty}{\frac{dp_{3}}{2\pi}{\ln {\left( 1-
e^{-p_{0}/T }\right)}}}, \label{f2}
\end{equation}
where $p_{0}=\sqrt{M^{2}+p_{3}^{2}}$ and the factor of 2
compensates for integrating over only positive values of $p_3$. It
is convenient to change variables from $p_{3}$ to $p_{0}$:
\begin{equation}
dp_{3}=\frac{p_{0}}{\sqrt{p_{0}^{2}-M^{2}}} dp_{0}. \label{f3}
\end{equation}
The limits of integration are changed from $0 \leq p_{3} < \infty$
to $M\leq p_{0} < \infty$.
We may also use the following representation for the logarithm
that appears in the integrand
\be
{\ln {\left( 1- e^{- p_{0}/T }\right)}}
 =-\sum_{j=1}^{\infty}\frac{e^{-j p_{0}/T}}{j},
\label{f4}
\ee
since $p_0$ is positive and $0\leq e^{- p_{0}/T}< 1$. Finally, we
obtain an expression for the total bosonic free energy just by
summing over the energy spectrum
\be
{\cal{F}}_{B}=-\frac{VT}{\pi}\sum_{n=1}^{\infty}
{\sum_{j=1}^{\infty}}\int_{M_{n}}^{\infty}{dp_{0}
\frac{p_{0}}{\sqrt{p_{0}^{2}-M_{n}^{2}}} {\left( \frac{e^{-j
p_{0}/T}}{j}\right)}} .
\label{f5}
\ee
The calculation of the fermionic contribution to the free energy
proceeds analogously, and we find the identical result with the
exception that there  is a factor of $(-1)^j$ inside the
summation. We can separate out the massless states from these
expressions and calculate their contribution explicitly. We know
that for resolution $K$ there are $(K-1)$ massless bosons and
$(K-1)$ massless fermions. Thus the contribution to the free energy
from massless states is
\begin{equation}
{\cal{F}}^{0}_{B}=-\frac{(K-1)\pi}{6} V T^2, \quad
{\cal{F}}^{0}_{F}=-\frac{(K-1)\pi}{12} VT^2 .
\label{f6}
\end{equation}
After doing the integral over $p_{0}$, we find for the total free
energy
\begin{equation}
\frac{{\cal{F}}(T,V)}{VT^2}=-\frac{(K-1)\pi}{4}
-\frac{2}{T\pi}\sum_{n=1}^{\infty}{\sum_{l=0}^{\infty}
}M_{n}\frac{K_{1}\left((2l+1)\frac{M_{n}}{T}\right)}{(2l+1)}.
\label{f9}
\end{equation}
The even terms, where $j=2l$ in the original sum, cancel between the fermion
and boson contributions. We have also factored out the temperature dependence
of the massless contribution and the volume dependence.

\begin{figure}[ht]
\begin{tabular}{cc}
\hspace*{-0.8cm}
\psfig{figure=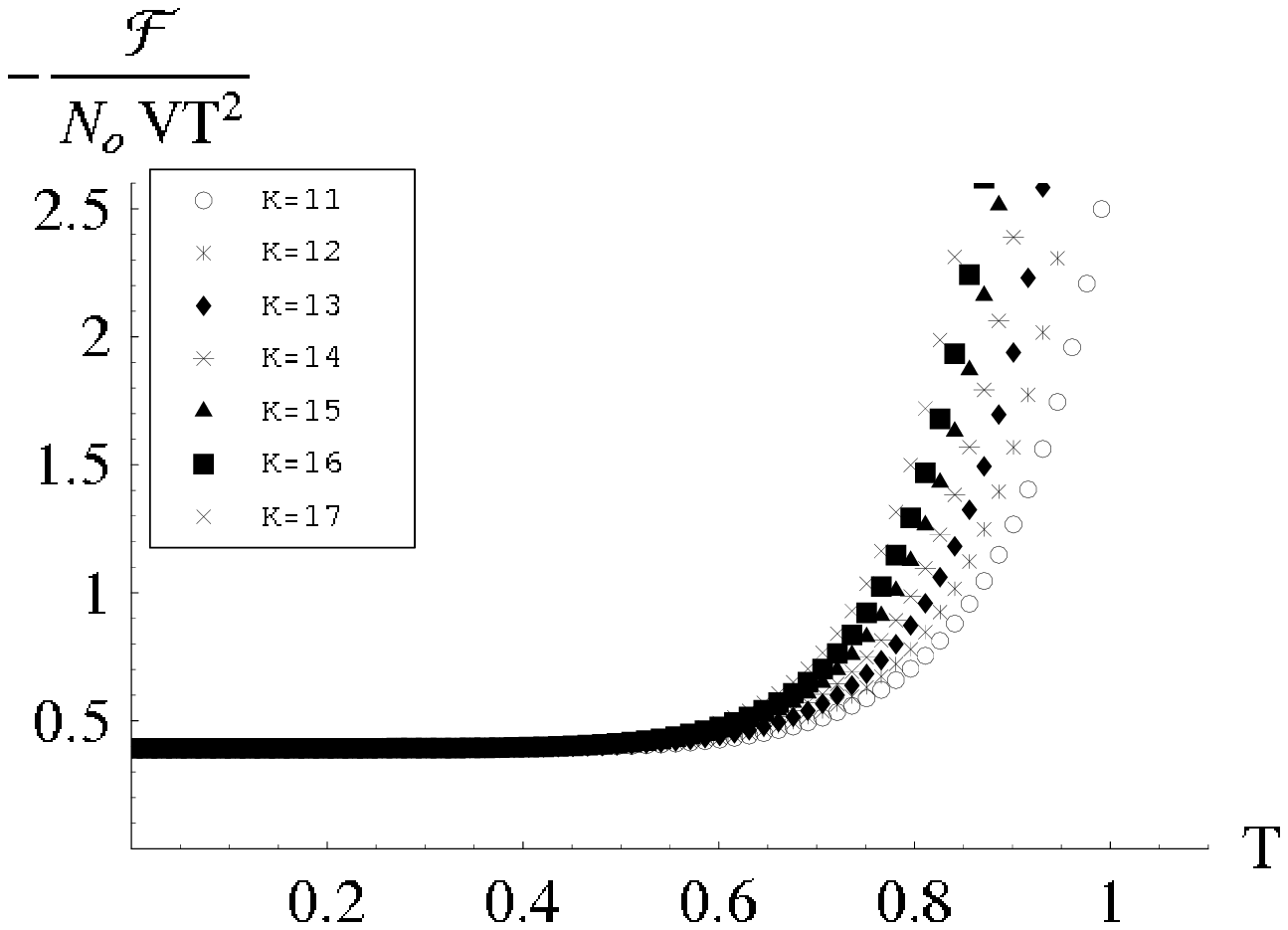,width=6.8cm,angle=0}&
\psfig{figure=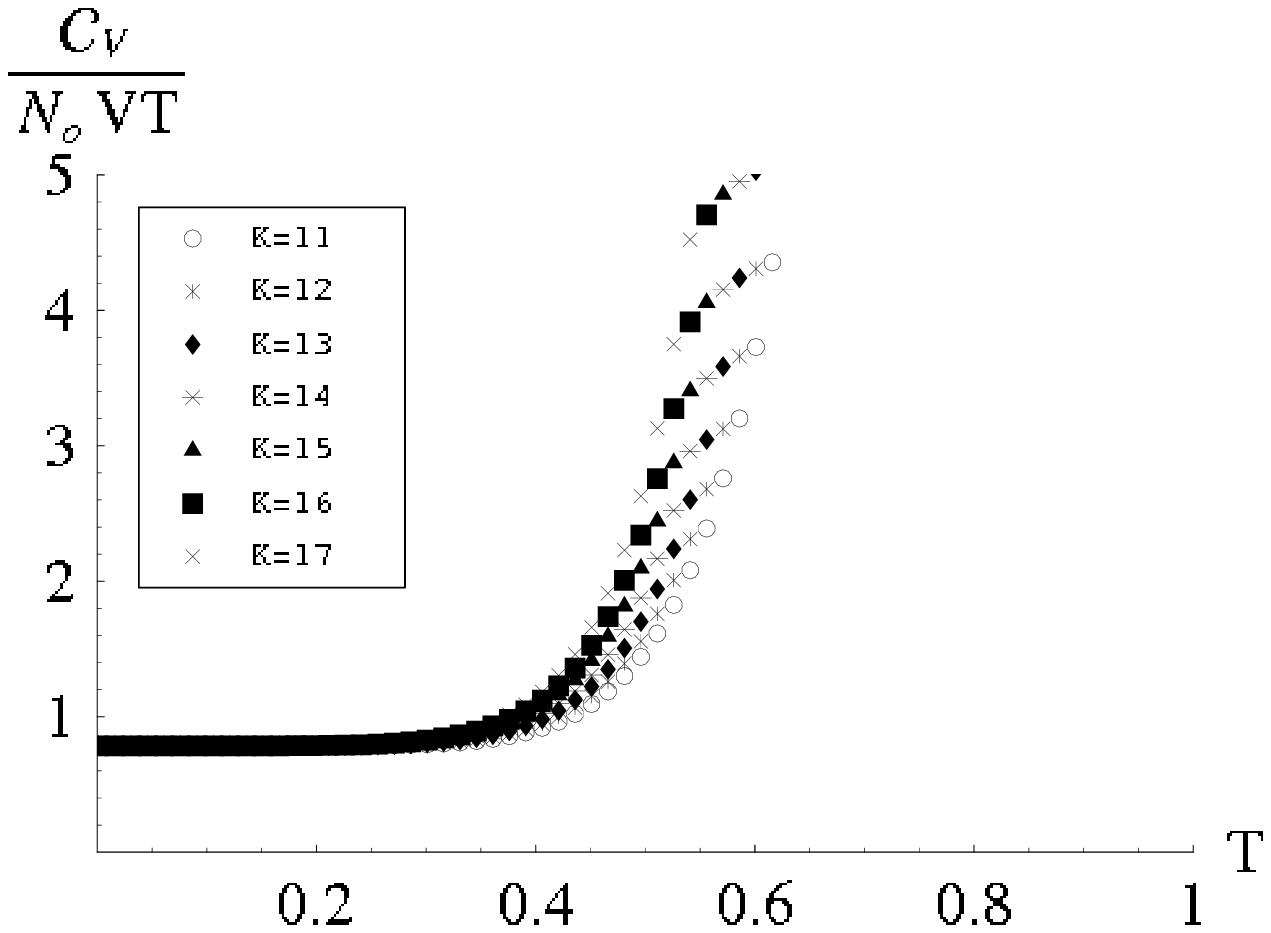,width=6.8cm,angle=0}\\
(a) & (b)
\end{tabular}
\caption{The free energy (a) and the heat capacity (b) as functions of
temperature for each resolution. Both functions are normalized to the
number of massless states, $N_0=2(K-1)$. The data points are a convenient
way to display the continuous functions calculated from fits to the CDF.}
\label{free}
\end{figure}

We can now rewrite the free energy in terms
of the density of states.  The
sums involving the Bessel function are cut off at a few terms; generally
$l_{\rm cut}\leq 2$ will be sufficient. We find
\begin{equation}
\frac{{\cal F}(T,V)}{VT^{2}}=-\frac{(K-1)\pi}{4}
-\frac{2}{\pi T} \int_{M_{\rm min}^{2}}^{\infty}dM^{2}
\rho(M^2)M \sum_{l=0}^{l_{\rm
cut}}\frac{K_1\left((2l+1)\frac{M}{T}\right)}{(2l+1)}.
\end{equation}

The free energy may now be used to calculate all the thermodynamic
functions. The internal energy  and heat capacity are given by,
\begin{equation}
{\cal E}(T,V)=T^2\left(\frac{\partial{\ln Z}}{\partial{T}}\right)_{\rm V},
\quad
{\cal C}_{\rm V}(T,V)=\left(\frac{\partial{{\cal
E}}}{\partial{T}}\right)_{\rm V} .
\label{f11}
\end{equation}

It is straightforward, given the density of states, to calculate the
thermodynamic functions.  Fig.~\ref{free}a shows the free energy, and
Fig.~\ref{free}b shows the heat capacity.  We expect the free energy to
diverge as $N_c^2$  and therefore must normalize our results
to extract a finite number. In most of the region below the Hagedorn
temperature  the thermodynamic functions are totally dominated by the
massless states. It therefore seems appropriate to normalize the
thermodynamic functions to the total number of massless bound states,
which is a function of the resolution and is $2(K-1)$.  Alternatively, we
could normalize by the number of states in any region. It
is conceivable that at very high resolutions, where the mass gap is
significantly less than one, that the massive states may make an important
contribution to the thermodynamics. In that case we would not choose to
normalize by the massless states.

%%%%%%%%%%%%%%%%%%%%%%%%%%%%%%%%%%%%%%%%%%%%%%%%%%%%%%%%%%%%%%%%%%%%%%%%%%%%
%
\section{Discussion}
\label{sec:discussion}
%%%%%%%%%%%%%%%%%%%%%%%%%%%%%%%%%%%%%%%%%%%%%%%%%%%%%%%%%%%%%%%%%%%%%%%%%%%%
%

The large-$N_c$ SDLCQ solution of ${\cal N}=(1,1)$ super Yang--Mills
theory  in 1+1 dimensions gives a set of non-interacting bound
states. From this set  of bound states it is in principle possible
to calculate the  thermodynamics of this  theory. Central to this
calculation is the calculation of the density of  states.
At resolutions $K=12$ and below, where we can  completely
diagonalize the Hamiltonian, we find that the entire cumulative
distribution function can
be fit with a single erf function. From the cumulative distribution
function, it is straightforward to calculate the density of states.
For $K$ larger than 12, it is  difficult to calculate the entire
spectrum; therefore, our calculations are  confined to a fixed range
of masses,
$M^2 \leq 100\,g^2  N_c/\pi$.   Using the known form of
the distribution, we only need to fit a  section of the cumulative
distribution function to get a very good fit to  the entire
distribution. We know analytically the total number of bound  states
at any resolution, and this information can also be used in
conjunction with a fit to a section of the distribution to
produce the  fit to the entire distribution.

The density of states
that are found by this  procedure grow sharply at small masses, then
level off and decrease at larger  masses. The peak of the density of
states grows as we increase the  resolution. Our understanding of
this behavior is that the cutoff of  the theory is forcing the density
of states to level off, turn over, and then decrease. The true behavior
of the density of states is reflected in the region  of the density of
states that is rapidly increasing, because it is this region that is
increasing in size with the resolution $K$.

It appears that the cumulative distribution function, and, therefore,
the  density of states, are growing exponentially with the mass. To
confirm this  and find the asymptotic values of this growth, we fit
the cumulative distribution function with an
exponential at each $K$. We  extrapolate these results to infinite
resolution to find the asymptotic behavior of the density of states.
The coefficient of the exponential growth is the reciprocal of the
Hagedorn temperature. We find that this temperature is
about $0.854\,\sqrt{g^{2} N_c/\pi}/k_B$. The
thermodynamic functions calculated from this data are expected to
produce  valid results up to a temperature that is around $T_{\rm
H}$.

It is now straightforward to calculate a standard set of
thermodynamic functions  from this density of states. The best
estimate of the thermodynamics is  obtained by using the exponential
fits to the density of states.  What we  see is that, for resolutions
up to $K=17$, all of the massive bound states are  well above the
Hagedorn temperature. The thermodynamics below $T_H$ is therefore
controlled by the $K-1$ massless boson bound states and $K-1$
massless  fermion bound states.

We can speculate on what will happen as the resolution goes to
infinity. We  have seen that the mass gap closes linearly with $1/K$.
So, for a resolution of order 100,  there will be massive bound states
below the Hagedorn temperature. This, of  course, assumes that the
estimate of the Hagedorn temperature is not changed  by the higher
resolution calculations.  We found, however, that the actual number of
massive bound states in a fixed mass  range may grow
slowly. For resolutions 11 to 17 we are able to  find excellent fits with
both exponential and linear growth as a function of  the resolution $K$
for masses with $M^2 \leq 100$. If the number of massive states grows only
linearly with $K$, the contribution to the thermodynamic functions below
$T_H$ might  become significant but not dominant.

These calculations indicate that ${\cal N}=(1,1)$ super Yang--Mills
theory  in 1+1 dimensions has a Hagedorn temperature of about one in
units of
$\sqrt{g^{2} N_c/\pi}/k_B$.  More generally, we found that  SDLCQ can
be used  to find interesting properties of finite-temperature
supersymmetric field  theories. The extension of this method to
theories with more supersymmetry  and in higher dimensions appears
to be straightforward but may be computationally challenging.

%%%%%%%%%%%%%%%%%%%%%%%%%%%%%%%%%%%%%%%%%%%%%%%%%%%%%%%%%%
\section*{Acknowledgments}
This work was supported in part by the U.S. Department of Energy
and by the Minnesota Supercomputing Institute.
%%%%%%%%%%%%%%%%%%%%%%%%%%%%%%%%%%%%%%%%%%%%%%%%%%%%%%%%%%

%
\end{document}